\documentclass[prl, aps, twocolumn, groupedaddress, superscriptaddress]{revtex4}

\usepackage{slashed}
\usepackage{graphicx}
\usepackage{subfigure}
\usepackage[usenames, dvipsnames]{color}
\usepackage{graphics}
\usepackage{hyperref}
\usepackage{bm}
\usepackage{amsfonts}

\begin{document}

\title{Topological Wigner Crystal of Half-Solitons in a Spinor BEC}

\author{H. Ter\c{c}as}
\email{htercas@gmail.com}
\author{D. D. Solnyshkov}
\author{G. Malpuech}
\affiliation{Institut Pascal, PHOTON-N2, Clermont Universit\'e, Blaise Pascal University, CNRS,24 Avenue des Landais, 63177 Aubi\`ere Cedex, France}

\begin{abstract}
We consider a one-dimensional gas of half-solitons in a spinor Bose-Einstein condensate. We calculate the topological interaction potential between the half-solitons.  Using a kinetic equation of the Vlasov-Boltzmann type, we model the coupled dynamics of the interacting solitons. We show that the dynamics of the system in the gaseous phase is marginally stable and spontaneously evolves towards a Wigner crystal. 
\end{abstract}
\maketitle

Since the early days of quantum mechanics, it is commonly accepted that the Wigner crystal is one of the most simple yet dramatic many-body effects. In the seminal work published in 1931 \cite{wigner}, Wigner showed that as a result of the competition between the long-ranged potential and kinetic energies, electrons spontaneously form a self-organized crystal at low densities, in a state that strongly differs from the Fermi gas. Experimental observations of this effect have been reported in carbon nano-tubes \cite{nanotubes}. The phenomenon has been discussed in ultracold Fermi gases with dipolar interactions \cite{dipolar, lakomy} and also investigated in systems with short-range interactions \cite{bloch, stefan}. The concept of Wigner crystal is used for the description of more exotic systems, such as holons (charge solitons) in the vicinity of a metal-Mott insulator transition \cite{krive}, and the ground-state properties of nuclear matter \cite{nuclear}.\par
In this Letter, we study the behavior of a one-dimensional gas of dark half-solitons (HSs). HSs are the elementary topological excitations of a one-dimensional spinor Bose-Einstein condensate with spin-anisotropic interactions \cite{Volovik}. We derive the interaction potential between HSs. We show that in some density range, the uncorrelated HS gas becomes unstable and undergoes Wigner crystallization, as a result of the competition between the fermionic statistics and interactions. Wigner crystal is usually associated with long-range interactions, such as the Coulomb interaction of electrons. Although interactions between HSs are short-ranged, we show that their specific form allows the formation of a quasi-ordered state exhibiting the quantum signature ($4k_F$-density correlations) of a usual Wigner crystal \cite{schulz}. We make use of a kinetic theory and perform numerical simulations to demonstrate the dynamical crystallization of a two-species HS gas. We compute the spectrum of the ordered phase and observe that it contains both acoustic and optical modes, showing that the optical gap is a function of the spin anisotropy. By estimating the density-density correlation within the Luttinger liquid theory, we demonstrate that the ordered state indeed corresponds to a topological version of the Wigner crystal. This ordered state could be used to model the low-energy features of topological excitations in spinor BECs in more general situations. For example, it makes possible the investigation of many-body effects in soliton tunneling, which is a spectacular yet not obvious quantum phenomenon in the nonlinear regime \cite{verde, dekel, assaf}. Additionally, there is a growing interest in the physics of HSs, mainly due to their spin textures, allowing them to behave  as magnetic charges in the presence of an effective magnetic field \cite{hivet}. Therefore, the study of both the ground-state and dynamical properties of an ensemble of HSs can be of a practical interest for the future development of magnetricity \cite{Bramwellnw} .

In what follows, we consider that HSs are traveling solutions of the 1D spinor Gross-Pitaevskii equation \cite{pitaevskii}
\begin{equation}
i\hbar \frac{\partial\psi_\sigma}{\partial t}=-\frac{\hbar^2}{2m}\nabla^2\psi_\sigma+\left(\alpha_1\vert \psi_\sigma\vert^2+\alpha_2 \vert \psi_{-\sigma}\vert^2\right)\psi_\sigma.
\label{GP}
\end{equation}
$m$ is the boson mass, $\alpha_1$ the interaction constant between particles having the same spin, and $\alpha_2$ the interaction constant between particles having opposite spins. When one soliton is present in a spin component and absent in the other, the corresponding object is called half soliton. The stability of HSs is ensured in the case of spin-anisotropic interactions, when $\vert \alpha_1\vert \gg \vert \alpha_2\vert $ which is typically realized in exciton-polariton BEC \cite{shelykh,Flayac}. Let us consider a HS formed in the $\sigma_+$ spin projection, for definiteness. In that case, the single-soliton solution traveling with speed $v$ is given by  
\begin{equation}
\psi_+(x)=\sqrt{n_0/2}\left[i\beta +\sqrt{1-\beta^2}\tanh\left(\sqrt{1-\beta^2}\frac{x}{\sqrt{2}\xi}\right)\right],
\label{soliton}
\end{equation} where $\beta=v/c_s$, $c_s\simeq \sqrt{\alpha_1 m/c_s}$ is the sound velocity and $\xi=\hbar/\sqrt{2m\alpha_1n_0}$ denotes the healing length. From mechanical arguments (see the supplemental material \cite{suppl} for details), we can derive the pseudo-potential associated with the interaction between the two $\sigma_+$ solitons \cite{frantzeskakis, konotop}
\begin{equation}
V_{\mathrm{int}}(x)=\frac{M_*}{2} c_s^2 \frac{1-\beta^2}{\mbox{sinh}^2\left( \frac{ \sqrt{2}\sqrt{1-\beta^2}z}{\xi}\right)},
\label{solinteract}
\end{equation}
where $M_*$ represents the effective mass of the solitons. In fact, it reduces to an effective potential for the case of almost black ($v~\sim 0$) collisions, which reads
\begin{equation}
U(x)=\frac{M_*}{2} c_s^2\mbox{cosech}^2\left( \frac{\sqrt{2} x}{\xi}\right),
\label{poteff}
\end{equation} 
In order to incorporate the interaction with the $\sigma_-$ spin projection, we have to take into account the spinor nature of the ground state of the condensate. In that case, the function of two HSs in different components located at positions $x_1$ and $x_2$, traveling with opposite speeds $v$ and $-v$ 
can be written using the center-of-mass and relative coordinates $\zeta=(x_1+x_2)/2$ and $\eta=(x_1-x_2)/2$ respectively \cite{santos}. Integrating over the soliton centroid $\zeta$ to compute the energy $E=\int \mathcal{E}~d\zeta$, with $\mathcal{E}$ denoting the energy density (see \cite{suppl} for details), we can obtain the following potential between almost black $\sigma_+$$-$$\sigma_-$ solitons
\begin{equation}
V(x)=\frac{M_*}{2}\vert \Lambda \vert  c_s^2\left[ \frac{\sqrt{2} 	x\cosh(\sqrt{2}x/\xi)-\sinh(\sqrt{2}x/\xi)}{\sinh^3(\sqrt{2}x/\xi)} \right],
\label{poteff2}
\end{equation}
where $\vert \Lambda \vert=\vert \alpha_2\vert /\alpha_1$ is the measure of the spin-anisotropy of the interactions. This potential is repulsive for attractive inter-spin interactions ($\alpha_2<0$), which is the case considered here. Qualitatively, each HS creates a density dip in the other component, and this dip repels the other HS because of the negative mass of the latter.

\begin{figure}[t]
\flushleft
\includegraphics[scale=0.9]{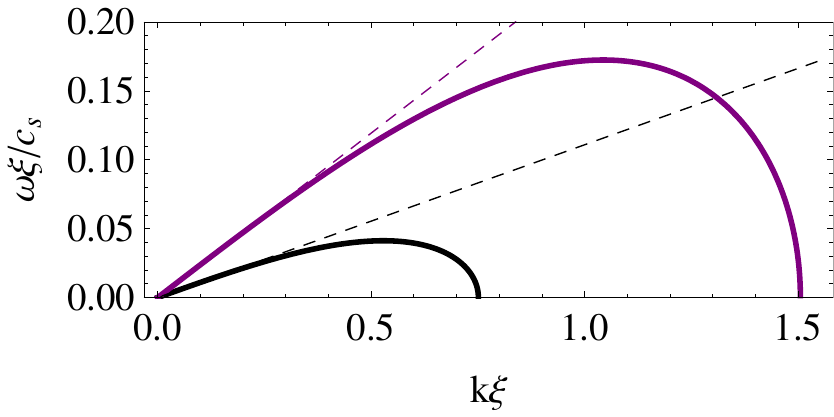}
\includegraphics[scale=0.9]{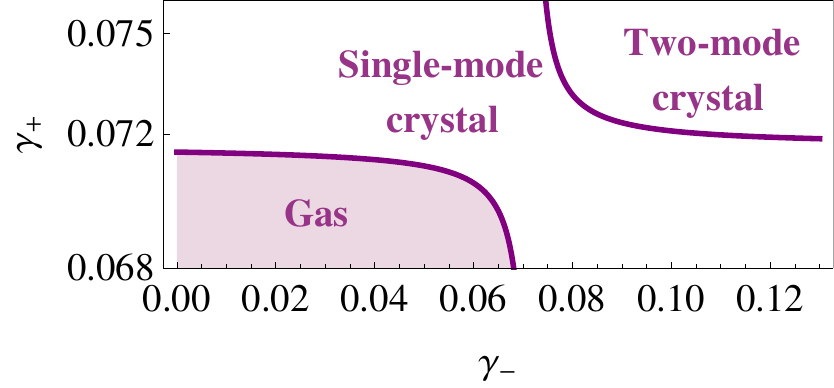}
\caption{(color online) Top: dispersion relation of a HS gas exhibiting dynamical instability, obtained for $\gamma_+=0.1$, $\gamma_-=0.08$ and $\vert \Lambda \vert =0.2$. The black (purple) full lines corresponds to $\omega_-$ ($\omega_+$) modes in Eq. (\ref{solkin3}). The dashed lines illustrate to the low-wavelength limit $\omega_\pm \simeq v_\pm k$. Bottom: phase diagram in the $(\gamma_+,\gamma_-)$ plane for $\Lambda=0.2$, depicting the gaseous, single-mode and two-mode crystalline phases discussed in the text.}
\label{solgasdisp}
\end{figure}


To describe the dynamics of a gas of HSs, we postulate that the phase-space distributions $f^\pm(x,v,t)$ are governed by following kinetic equation of the Vlasov type
\begin{equation}
\frac{df^\pm}{dt}\equiv \frac{\partial f^\pm}{\partial t}+v\frac{\partial f^\pm}{\partial x}+\dot v^\pm\frac{\partial f^\pm}{\partial v}=0,
\label{solkin1}
\end{equation} 
where the collision integral is neglected assuming that only elastic processes are involved in the system. The acceleration term can be simply given by $\dot v^\pm=-1/M_* \partial U^\pm_\mathrm{eff}/\partial x$, where $U^\pm_\mathrm{eff}(x)$ is given in terms of the topological potentials (\ref{poteff}) and (\ref{poteff2})
\begin{equation}
\begin{array}{c}
\displaystyle{U^\pm_\mathrm{eff}(x)=\int_{-\infty}^{\infty}\int_{-\infty}^{\infty}U(x-x')f^\pm(x',v)dx'dv}\\
\displaystyle{+\int_{-\infty}^{\infty}\int_{-\infty}^{\infty}V(x-x')f^\mp(x',v)dx'dv}
\end{array}.
\end{equation}
Equation (\ref{solkin1}) generalizes the continuity equation presented in Ref. \cite{el} by the introduction of the soliton interaction potential. We now linearize the system around its equilibrium, $f^\pm=f_0^\pm+\delta f^\pm$ to obtain the following dispersion relation \cite{suppl} for the excitations of the soliton gas
\begin{equation}
(1-I_U^+)(1-I_U^-)-I_V^-I_V^+=0,
\label{solkin2}
\end{equation}
where the integrals $I_U^\pm$ and $I_V^\pm$ have the form 
\begin{equation}
I_X^\pm(k,\omega)=\frac{k^2 \tilde X(k)}{M_*}\int_{-\infty}^\infty \frac{f_0^\pm}{(\omega-kv)^2}dv,
\label{solkin3}
\end{equation}
with $\tilde X(k)$ representing the Fourier transform of the potentials $U$ and $V$. It is known that the excitations in one-dimensional Bose gases below the critical velocity $c_s$ follow a fermionic statistics, as established in the famous Lieb-Liniger theory \cite{lieb1, lieb2}. Actually, this result is quite easy to understand by simply looking at the phase of the two-soliton wave function (1) of \cite{suppl}: exchanging two solitons located at positions $x_1$ and $x_2$, we obtain an overall phase shift of $\pi$ in agreement with fermionic statistics. Therefore, the equilibrium configuration is that of a one-dimensional Fermi gas $f_0(x,v)^\pm=N_0^\pm/2v_F^\pm\Theta(v_F^\pm-\vert v\vert),$
where $v_F^\pm=\pi \hbar N_0^\pm/ M_*$ is the one-dimensional Fermi velocity and $N_0^\pm$ is the soliton density. In that case, the excitation spectrum above the equilibrium configuration contains two branches $\omega_\pm(k)$ displaying acoustic behavior in the long wavelength limit $k\rightarrow 0$ as $\omega_\pm\approx v_\pm k$, where the velocities are given by cumbersome expressions of $N_0^\pm$ \cite{suppl}. In the special case of a symmetric gas $N_0^+=N_0^-=N_0$, we can obtain 
\begin{equation}
v_\pm=\frac{v_F}{2\sqrt{2}\pi\gamma}\sqrt{8\pi^2\gamma^2-\sqrt{2}\gamma(4\pm\vert\Lambda\vert)},
\label{estimate1}
\end{equation}
with $\gamma=N_0\xi$ representing the dimensionless soliton concentration, an analogue of the Wigner-Seitz parameter. We have here used the relation between the Fermi velocity of the soliton gas and the sound velocity of the condensate, $v_F/c_s=\sqrt{2}\pi \gamma$, which has the meaning of a dimensionless interaction parameter. 

Due to the competition between the statistical pressure and the interactions between the solitons, the dispersion relation (\ref{solkin2}) encodes very interesting features, as illustrated in Fig. (\ref{solgasdisp}). If the soliton gas is dilute enough so that the concentration parameter lies below the critical value $\gamma^*\simeq 0.07$, the system does not exhibit any ordering, corresponding to a gaseous state. Instead, for $\gamma>\gamma^*$, the solitons start to perform periodic oscillations up to a critical value of the wavevector $k^*=k^*(\gamma)$. At this point, the frequency softens towards zero and the system undergoes crystallization with the lattice constant given by $d=2\pi/k^*$. The onset of instability is expected at higher $k^*$ for higher values of $\gamma$, therefore corresponding to tighter lattices. When the interactions dominate and the distance becomes comparable with the characteristic scale of these interactions, the system forms a regular lattice. The ratio of the Fermi velocity to the interaction potential (\ref{solinteract}) provides the following magnitude estimate $\left\vert v_F^2/U \right\vert \sim v_F^2/c_s^2\sim\gamma^2$. For the case of an asymmetric mixture, a similar analysis allows us to conclude about the existence of different phases: a gaseous phase ($v_-=v_+=0$), defined by the region $\gamma_+<(16-\vert\Lambda \vert^2-16\sqrt{2}\pi^2\gamma_-)/[16\pi^2(\sqrt{2}\pi^2\gamma_-)]$, and two ordered  phases, sustaining single-mode ($v_-=0, v_+>0$) and a two-mode ($v_-,v_+>0$) oscillations. The latter two situations dynamically evolve into a crystalline phase sustaining one and two phonon modes, respectively. We illustrate the dispersion relation and the onset of instability associated to mode softening (Fig. (\ref{solgasdisp})a), together with the phase diagram (Fig. (\ref{solgasdisp})b), which is zoomed in the relevant region of the plane ($\gamma_+,\gamma_-$) where the transitions between the different regimes occur.\par
\begin{figure}[t!]
\flushleft
\includegraphics[scale=0.90]{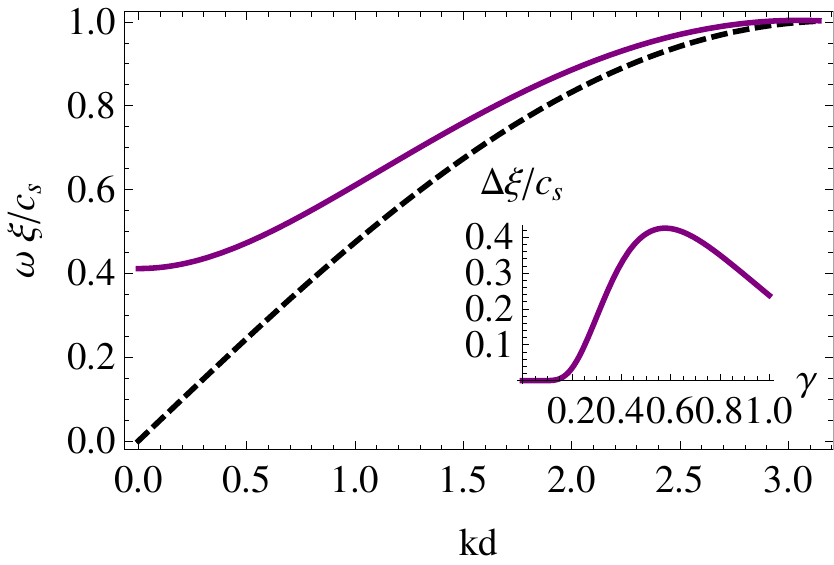}
\caption{(color online) Spectrum of a symmetric crystal with $\gamma_+=\gamma_-=0.5$, corresponding to the half-filling configuration. Acoustic (black) and optic (purple) phonon modes. The inset represents the gap frequency $\Delta$ as a function of the parameter $\gamma$.}
\label{crystaldisp}
\end{figure}

 
In what follows, we analyze the two-mode ordered state and show that it indeed corresponds to a crystal exhibiting the same quantum properties as a Wigner crystal. The configuration of minimum potential energy is obtained for a chain of alternating $\sigma_+ -\sigma_-$ solitons with lattice constant $d$ and quantum oscillations are expected to lead only to low-amplitude oscillations around the equilibrium distance. For definiteness, we consider the half-filling configuration $d=1/N_0$. In the harmonic approximation, the corresponding Hamiltonian reads
\begin{equation}
\begin{array}{c}
\displaystyle{H=\sum_\ell \left( \frac{p_{u,\ell}^2}{2M_*}+\frac{p_{v,\ell}^2}{2M_*}\right)+\frac{1}{4}\sum_{\ell,\ell'} U''(\ell d)(u_\ell-u_{\ell+2\ell'})^2}\\
\displaystyle{+\frac{1}{4}\sum_{\ell,\ell'} V''(\ell d)(u_\ell-v_{\ell+\ell'})^2,}
\end{array}
\label{crystalham}
\end{equation}
where $u_\ell$ ($v_\ell$) is the deviation of the $\sigma_+$($\sigma_-$)-soliton at site $\ell$ from its equilibrium position. The diagonalization of (\ref{crystalham}) with the help of Hamilton equations leads to two modes, an acoustic mode $\omega_1$ and a gapped optical mode $\omega_2$. In the long-wavelength limit they are given by $\omega_1 \approx u_1 k$ and $\omega_2^2=\Delta^2+u_2^2k^2$, where the velocities $u_1$ and $u_2$ are functions of the ratio $d/\xi=1/\gamma$ \cite{suppl}. Within the first neighbor approximation, we obtain
\begin{equation}
u_{1,2} \simeq d \sqrt{\frac{ U''(d)}{M_*}\pm \frac{ V''(d)}{4M_*}}, \quad \Delta \simeq 2 \sqrt{\frac{V''(d)}{M_*}}.
\end{equation}
The {\it plasma} frequency $\Delta$ is a feature that distinguishes the crystal from the Fermi gas, and is a function of the spin-anisotropy $\vert \Lambda\vert$, $\Delta\sim \sqrt{\vert \Lambda \vert}$. In Fig. (\ref{crystaldisp}), we plot the phonon modes $\omega_{1,2}$ in the first-neighbor approximation for the case of a symmetric crystal.\par
\begin{figure}[ht!]
\centering
\includegraphics[scale=0.9]{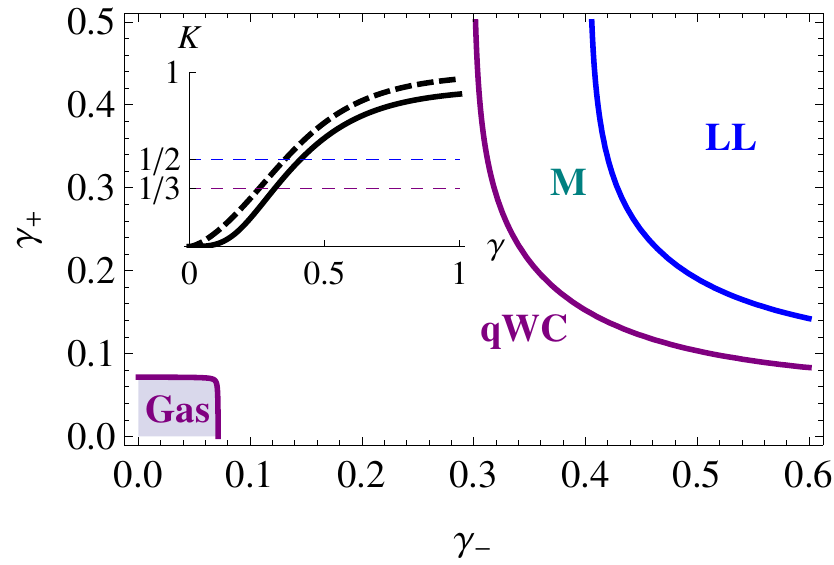}
\caption{(color online) Phase diagram of the ordered phase described by the Hamiltonian (\ref{crystalham}). The contours $K=1/2$ (blue line) and $K=1/3$ (purple line)  delimit the Luttinger Liquid (LL), the topological quasi-Wigner crystal (qWC) and the mixture (M) phases. Inset: Cut the Luttinger parameter $K$ along  $\gamma_+=\gamma_-=\gamma$ for single (dashed) and two-species (solid) crystal. The small dashed region represents a zoom-out of the gas phase of Fig. (\ref{solgasdisp}).}
\label{crystalcorr}
\end{figure}

\begin{figure}[h!]
\centering
\includegraphics[scale=0.25]{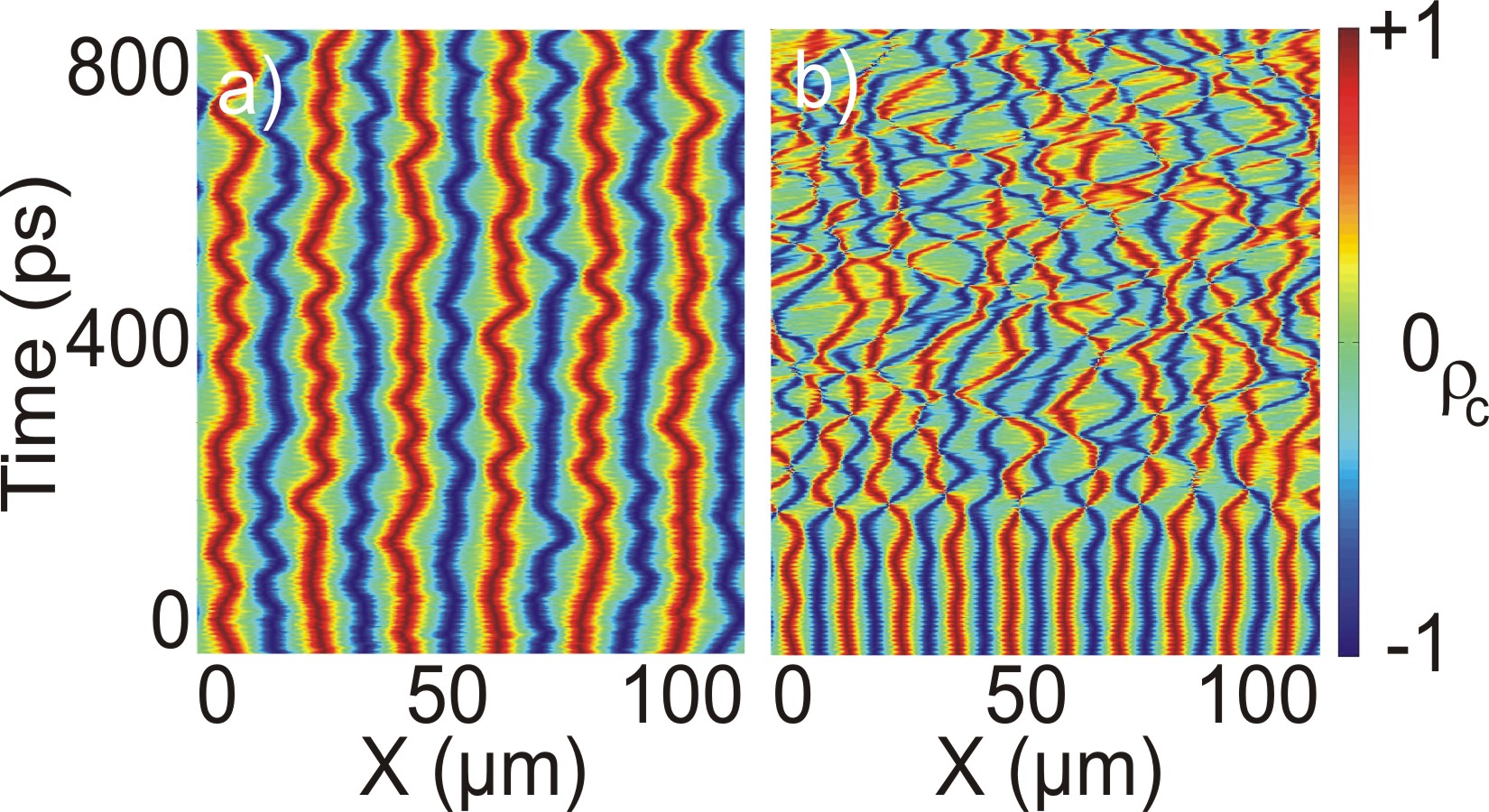}
\caption{(color online) Circular polarization degree of the condensate as a function of time showing (a) a stable Wigner crystal ($\gamma_+=\gamma_-\approx 0.25$),  and (b) "melting" of such crystal into a liquid-like state, ($\gamma_+=\gamma_-\approx 0.5$). In both situations, we took $\vert \Lambda \vert =0.1$.}
\label{solsimul}
\end{figure}

In order to investigate the quantum features of the crystalline phase, we compute the density correlations. Due to the short range character of the interactions, the Luttinger liquid theory \cite{haldane} can accurately describe the physical properties of the system. For a model with $SU(2)$ spin symmetry such as Hamiltonian (\ref{crystalham}) (or, more precisely, its second-quantized version), bosonization predicts the decay of the density correlation function as follows \cite{schulz2}
\begin{equation}
\langle n_xn_0\rangle \simeq - \frac{K}{\pi x^2}+\frac{A_2}{x^{1+K}}\cos(2k_Fx)+\frac{A_4}{x^{4K}}\cos(4k_Fx),
\label{corr}
\end{equation}  
where $A_2$ and $A_4$ are some constants. The Luttinger parameter \cite{haldane, sirker} $K=\sqrt{K_U^3K_V}/(K_UK_V+\vert\Lambda \vert)$ can be given in terms of the single-spin parameters \cite{suppl}  
\begin{equation}
K^{-1}_X=1+\frac{2 }{\pi \hbar M_* v_F}\sum_\ell \xi X(\ell d)\left[1-\cos\left(2k_F \ell d\right)\right],
\end{equation}
with $X=U,V$. Apart from the $x^{-2}$ dependence, which is familiar from the Fermi liquid theory, the properties of the system are universally defined in terms of $K$. Therefore, for $K>1/2$, the $2k_F$ Friedel-like oscillations dominate, which is typical of a Luttinger liquid (LL). On the contrary, for $K<1/3$, $4k_F$ quantum fluctuations dominate the system, leading to a modulation at the average distance $d=1/N_0$ between the solitons. This corresponds to a quasi-Wigner crystal (qWC) state, where the (quasi) long-range order is due to quantum fluctuations, being favored in the low density limit (notice that the interaction energy scales as $c_s/v_F\sim 1/\gamma$). In the intermediary region $1/3<K<1/2$, there is a mixture (M) between the two previous phases. An interesting feature distinguishing qWC stated from the Wigner crystal of electrons is the topological nature of the effective potentials, whose short range is defined by the condensate's healing length $\xi$. This suggests that this state can be regarded as a topological Wigner crystal. Yet another difference is related with the absence of the logarithmic divergence of the Coulomb potential $\sim 1/x$ leading to a very slow decay in the $4k_F$ correlation (much slower than any power law \cite{schulz}). Excitations about the qWC equilibrium corresponds to the topological equivalent of the so-called charge density wave, occurring for $K<1/4$ and characterized by commensurate condition of the filling factor $\gamma=1/\nu$, where $\nu$ is an integer. In Fig. (\ref{crystalcorr}) we illustrate the relevant phases occurring in the system. We can observe that it exhibits the same phases as in electronic systems, except for the gaseous phase occurring for very dilute systems. This stems in the short-range nature of the topological interactions between the HSs. We performed numerical simulations of the Gross-Pitaevskii equation (\ref{GP}) to illustrate the existence of both crystallized and liquid phases. In these simulations, the initial configuration was a condensate with a Fermi distribution of HSs. Fig. (\ref{solsimul}) depicts the circular polarization degree $\rho_c=(n_+-n_-)/(n_++n_-)$, which is the most suitable quantity to track the dynamics of HSs in BECs. In one case (panel a), the regular pattern corresponding to Wigner crystal is conserved while HSs are oscillating around their equilibrium positions; in the other case (panel b), the pattern changes dramatically and the ordering is lost, as it is typical of a liquid-like state.\par Notice that the present analysis is valid in zero temperature limit with no disorder. Beyond the obvious modification of the distribution, it is known that both temperature and disorder can melt the crystal state \cite{nagara}. In analogy with the criteria used in one-component plasmas \cite{march}, the melting occurs for $\Gamma\equiv E_\mathrm{int}/{E_\mathrm{kin}}\sim 170$. Therefore, we should expect the crystal to be robust for $\hbar \Delta\gg E_\mathrm{int},E_\mathrm{dis}$, where $E_\mathrm{dis}$ is the disorder amplitude. A more precise estimation  is out of the scope of the present work.

In conclusion, we have investigated the crystallization of a gas of HSs in a spinor Bose-Einstein condensate. Starting with a kinetic equation to describe the coupled dynamics of two-spin species mixture in Bose-Einstein condensate, we have shown that the system sustains unstable oscillations about the Fermi gas equilibrium and spontaneously undergoes crystallization, as a consequence of the competition between the interactions and the Fermi statistics. For the ordered phase, we computed the spectrum of collective excitations, obtaining two phonon modes (corresponding to the acoustic and optical branches). The analysis of the low-energy fluctuations based on the Luttinger liquid theory revealed that the ordered phase indeed corresponds to a Wigner crystal, where the $4k_F$ correlations are induced by the topological effective potential between the solitons. \par

We would like to thank H. Flayac for discussions. This work was supported by ANR ``QUANDYDE", FP7 ITN  ``Spin-Optronics"(237252), and the FP7 IRSES ``POLAPHEN" (246912).

\end{document}